\documentclass[conference]{IEEEtran}
\IEEEoverridecommandlockouts
\usepackage{cite}
\usepackage{amsmath,amssymb,amsfonts}
\usepackage{algorithmic}
\usepackage{graphicx}
\usepackage{textcomp}
\usepackage{xcolor}
\usepackage{url}
\usepackage{multirow}
\usepackage{subfigure}
\def\BibTeX{{\rm B\kern-.05em{\sc i\kern-.025em b}\kern-.08em
    T\kern-.1667em\lower.7ex\hbox{E}\kern-.125emX}}

\begin{document}

\title{SELIC: Semantic-Enhanced Learned Image Compression via High-Level Textual Guidance}

\author{
\IEEEauthorblockN{Haisheng Fu \\
Simon Fraser University\\
haisheng\_fu@sfu.ca}
\and
\IEEEauthorblockN{Jie Liang \\Simon Fraser University\\
jie\_liang@sfu.ca}
\and
\IEEEauthorblockN{Zhenman Fang \\
Simon Fraser University\\
zhenman@sfu.ca}
\and
\IEEEauthorblockN{Jingning Han \\
Google\\
jingning@google.com}
\thanks{This work was supported by the Natural Sciences and Engineering Research Council of Canada (RGPIN-2020-04525, RGPIN-2019-04613, DGECR-2019-00120, ALLRP-552042-2020), Google Chrome University Research Program, MITACS Elevate Postdoc Program IT42556, CFI John R. Evans Leaders Fund.}
}

\maketitle

\begin{abstract}
Learned image compression (LIC) techniques have achieved remarkable progress; however, effectively integrating high-level semantic information remains challenging. In this work, we present a \underline{S}emantic-\underline{E}nhanced \underline{L}earned \underline{I}mage \underline{C}ompression framework, termed \textbf{SELIC}, which leverages high-level textual guidance to improve rate-distortion performance. Specifically, \textbf{SELIC} employs a text encoder to extract rich semantic descriptions from the input image. These textual features are transformed into fixed-dimension tensors and seamlessly fused with the image-derived latent representation. By embedding the \textbf{SELIC} tensor directly into the compression pipeline, our approach enriches the bitstream without requiring additional inputs at the decoder, thereby maintaining fast and efficient decoding. Extensive experiments on benchmark datasets (e.g., Kodak) demonstrate that integrating semantic information substantially enhances compression quality. Our \textbf{SELIC}-guided method outperforms a baseline LIC model without semantic integration by approximately 0.1-0.15 dB across a wide range of bit rates in PSNR and achieves a 4.9\% BD-rate improvement over VVC. Moreover, this improvement comes with minimal computational overhead, making the proposed \textbf{SELIC} framework a practical solution for advanced image compression applications.
\end{abstract}

\section{Introduction}
\label{sec:intro}

In recent years, deep learning tools have been extensively applied to the field of image compression, achieving remarkable advancements that surpass traditional standards such as JPEG \cite{JPEG}, JPEG 2000 \cite{JPEG2000}, BPG (intra-coding of H.265/HEVC) \cite{BPG}, and H.266/VVC \cite{VVC} in both objective and subjective metrics. Traditional image compression techniques typically consist of key components such as linear transform functions, quantization modules, and entropy coding. Similarly, end-to-end learned image compression frameworks follow a comparable pipeline but leverage learnable parameters to replace and enhance these modules.

Early learned image compression models primarily focused on enhancing coding performance by introducing various advanced neural network architectures, including residual blocks \cite{Chen_TNNLS_2022, FU_2020, Asymmetric_Fu}, self-attention mechanisms \cite{chen2021, Li_Nonlocal_entropy}, invertible structures \cite{xie2021enhanced}, transformer-based blocks \cite{zhu2022transformerbased, Zou_2022_CVPR}, and wavelet-based blocks \cite{Ma_PAMI, Fu_EECV}. These modules enable neural networks to extract effective and efficient latent representations while reducing the redundancy in the input image.

Estimating a powerful and efficient entropy model is a critical topic in image compression. In \cite{Variational}, a hyperprior network is proposed to estimate the conditional probabilities of the latent representations. The hyper encoder extracts hyperpriors from the latents, which are encoded as side information and transmitted to a hyper decoder. The reconstructed hyperpriors enable the estimation of latent conditional probabilities, making the entropy model adaptive to both image content and spatial variations. This approach assumes a zero-mean Gaussian scale mixture (GSM) model for the latents, achieving superior performance compared to BPG (4:4:4). Subsequently, \cite{Joint} extends this method by employing a non-zero-mean Gaussian mixture model (GMM). Furthermore, the autoregressive context model introduced in \cite{pixelCNN} is utilized to refine latent probability estimation by leveraging both hyperpriors and spatial context.

Dynamic bitrate allocation is also an essential technique for improving image compression performance. It involves allocating more bits to important regions while assigning fewer bits to flat areas. Most previous methods \cite{Limu_conf, Asymmetric_Fu} achieve bitrate allocation by introducing importance maps. However, these dynamic maps are learned by the network and do not represent the high-level semantic information conveyed by the image.
A critical aspect of learned image compression is the integration of semantic information to enhance compression efficiency. While existing methods excel in capturing low-level image features, they frequently overlook the potential of high-level semantic guidance. Incorporating semantic information can provide a more nuanced understanding of image content, enabling more intelligent compression strategies that prioritize essential elements within an image.

In this paper, we introduce a novel image compression framework that leverages high-level semantic information extracted from images to guide the compression process. Our approach involves using a text encoder to translate image content into rich textual descriptions that encapsulate the image's key semantics. This textual information is then transformed into a fixed tensor vector, which is integrated with the encoded image bitstream through various fusion techniques. By embedding semantic vectors directly into the compression pipeline, our method enhances the bitstream without necessitating additional inputs during decoding, thereby maintaining efficient decoding speeds.

The main contributions of this paper can be summarized as follows:

\begin{itemize}
    \item We introduce a semantic-guided compression framework that leverages a pretrained text encoder to extract high-level semantic information from the input image. By embedding these semantic cues directly into the latent domain, our method preserves critical content and contextual nuances throughout the compression pipeline.

    \item We propose an image-text fusion module employing a channel-concatenation-based fusion strategy to effectively integrate the extracted semantic features with the image’s latent representation. Compared to conventional element-wise addition or multiplication, this approach provides a more robust and flexible mechanism for aligning semantic and visual signals, thereby enhancing compression performance with minimal computational overhead.

    \item Our framework avoids the need for external information at the decoder, as the semantic guidance is fully embedded in the transmitted bitstream. This not only streamlines and accelerates the decoding process but also delivers superior rate-distortion efficiency. Experimental results show that our method outperforms recent learned image compression (LIC) approaches, achieving a better trade-off among coding performance, decoding time, and model complexity.
\end{itemize}

Extensive experiments on benchmark datasets (e.g., Kodak and Tecnick) demonstrate that incorporating semantic information leads to notable improvements in PSNR. Our findings confirm that semantic-aware integration within the compression pipeline produces more faithful reconstructions and enhances visual fidelity across a wide range of bit rates.

\section{Related Work}
\subsection{Learned Image Compression}

Learned image compression methods have achieved superior rate-distortion performance in terms of PSNR and MS-SSIM metrics. These methods mainly improve coding performance through two ways. The first way is that different neural networks are provided to enhance encoder and decoder's learning abilites and better reduce the correlation of the latent representation. For example, different attention mechanisms \cite{cheng2020, GLLMM, Liu_2023_CVPR} are introduced to extract compact and efficient latent representations. Also, the transformer-based modules \cite{Liu_2023_CVPR, zhu2022transformerbased} are proposed to help the networks to extract global information of input image and better improve compression efficiency. 

The second approach involves estimating a powerful auto-regressive entropy model for latent representations. For instance, the GMM~\cite{cheng2020} and GLLMM~\cite{GLLMM} were proposed to accurately model the probability distribution of complex image regions, significantly reducing bit rates. Although these entropy models enhance compression performance, they introduce considerable decoding latency. To address this issue, parallelizable auto-regressive entropy modules, such as the checkerboard entropy model and the channel-wise auto-regressive entropy model (ChARM), have been proposed. These methods aim to accelerate the decoding process while maintaining coding performance as much as possible.

Achieving a good trade-off between coding performance and decoding speed is a crucial challenge in image compression tasks. For instance, knowledge distillation methods~\cite{Li_KD_2022, Asymmetric_Fu} have been proposed to reduce the complexity of decoding networks while preserving rate-distortion performance as much as possible.

\subsection{Text-Conditioned Learned Image Compression}

Early attempts at text-guided image compression, such as Text \& Sketch (PICS)~\cite{lei2023text+sketch}, combined caption information and sketch-based structures within diffusion models. While this approach achieved very low bitrates and preserved high-level semantics, it often produced images that differed significantly from the original content, as it prioritized semantic fidelity over pixel accuracy.

Later works like MISC~\cite{li2024misc} integrated both full-resolution images and text-driven diffusion refinements. Although MISC improved perceptual quality, it still used separate processing streams for text and image inputs. This separation, along with sparse textual encoding, made it difficult to maintain fine visual details (e.g., high PSNR) because semantic abstractions and low-level image signals were not fully aligned.

In contrast, our approach embeds textual semantics directly into the learned compression pipeline at the latent level. By fusing text and image representations before quantization, we remove the need for separate text-image pathways. This unified process retains key semantic cues throughout compression and reconstruction without relying on diffusion-based methods. As a result, we achieve strong semantic guidance while maintaining or even improving pixel-level fidelity, demonstrating a more direct and effective way to incorporate textual information into learned image compression.

\section{The Proposed Image Compression Framework}

\begin{figure*}[!thp] 
\centering 
\includegraphics[scale=0.5]{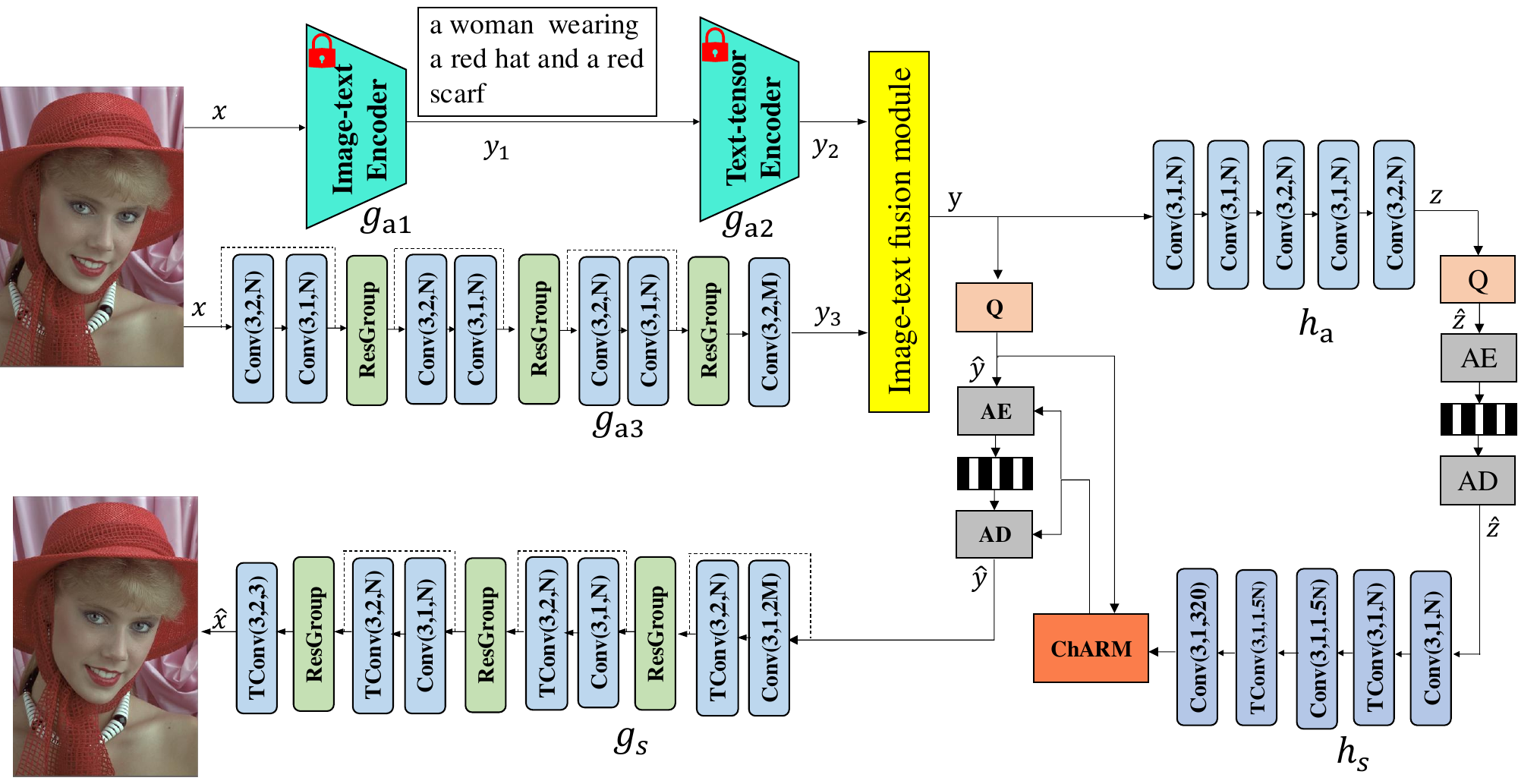} 
\caption{The detailed architecture of the proposed framework. \texttt{Conv(3, s, n)} denotes a convolutional layer with a $3 \times 3$ kernel size, stride $s$, and $n$ filters. \texttt{TConv(3, s, n)} represents a transposed convolution layer. Dashed shortcut connections represent changes in tensor size. \texttt{AE} and \texttt{AD} stand for Arithmetic Encoder and Arithmetic Decoder, respectively. The dotted lines represent the shortcut connection when size is change, as in \cite{resblock,cheng2020}.} 
\label{whole_networkstructure} 
\end{figure*}

Our proposed framework integrates high-level semantic guidance into a learned image compression pipeline, as illustrated in Fig.~\ref{whole_networkstructure}. It comprises a main image encoder $g_{a3}$, a main image decoder $g_{s}$, a semantic image-to-text encoder (BLIP \cite{BLIP}) $g_{a1}$, a text-to-tensor encoder (BERT \cite{BERT}) $g_{a2}$, a text-image latent fusion module, and hyperprior networks $h_{a}$ and $h_{s}$. Unlike conventional methods that rely primarily on low-level visual signals, our method enriches the latent representation with contextual semantic information derived from a pretrained text model.

The process unfolds as follows. Given an input image $x$, the semantic branch first converts it into a textual description via $g_{a1}$. This textual output is then encoded by $g_{a2}$ into a compact semantic feature vector of fixed dimension named $y_{2}$. In parallel, the main image encoder $g_{a3}$ processes the input image $x$ through three residual blocks \cite{resblock} and three downsampling steps to produce an image-derived latent $y_{3}$. The tensors $y_{2}$ and $y_{3}$ are then passed to the image-text fusion module to obtain $y$, as described in Sec.~\ref{fusion_model}.

It is important to note that both $g_{a1}$ (BLIP) and $g_{a2}$ (BERT) are directly taken from publicly available pretrained models without updating their parameters during training. Since these models have been extensively trained on large-scale datasets, they provide robust and semantically rich textual features, ensuring that our semantic guidance is of high quality without adding significant training overhead.

As in \cite{channel,Liu_2023_CVPR}, we employ a channel-wise entropy coding (ChARM) approach for the latent representation $y$, as indicated in Fig.~\ref{whole_networkstructure}. The ChARM structure remains consistent with \cite{Fu_DCC}. To enhance entropy coding performance, the hyperprior networks $h_a$ and $h_s$ encode and decode the hyperprior information $z$ for $y$, as shown in Fig.~\ref{whole_networkstructure}.

\subsection{Text-Image Latent Fusion Module}
\label{fusion_model}

\begin{figure*}[!thp] 
\centering 
\includegraphics[scale=0.55]{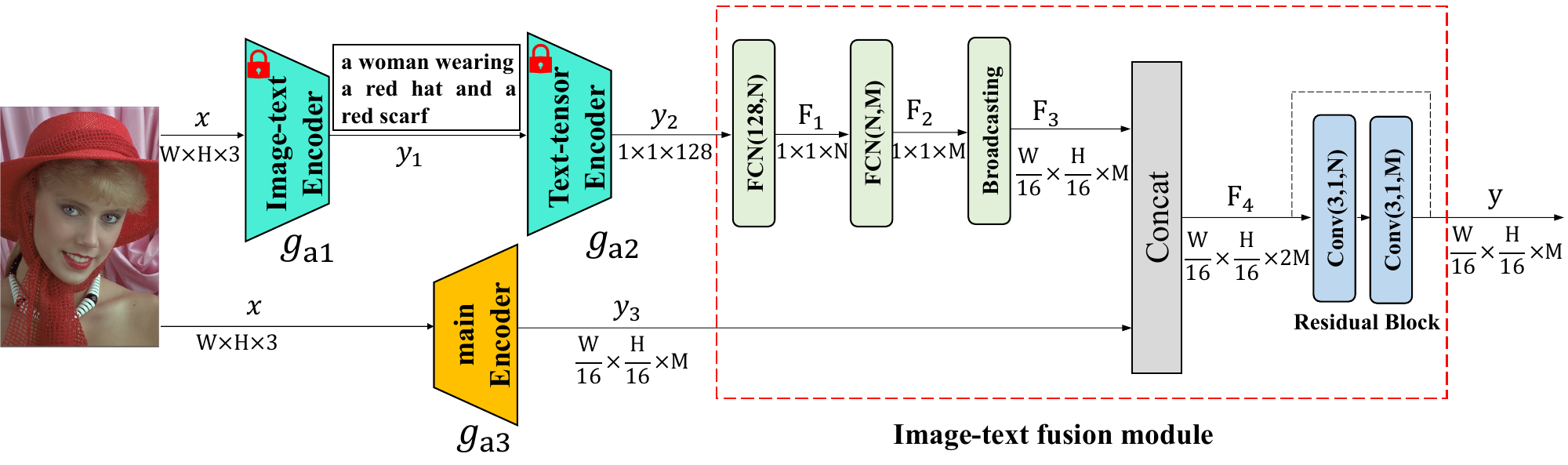} 
\caption{Processing details of the text-image fusion module. Each \texttt{FCN(N, M)} denotes a fully connected layer mapping from $N$ input units to $M$ output units.} 
\label{fusion_module}
\end{figure*}

As shown in Fig.~\ref{fusion_module}, the semantic representation $y_{2}$, initially a single vector of dimension $M$, is broadcast across the spatial dimensions to form a $(H/16) \times (W/16) \times M$ tensor, aligning high-level semantic features with the image grid. In parallel, the main image encoder $g_{a3}$ transforms $x$ into a latent $y_{3}$ of identical spatial size and $M$ channels. Ensuring both $y_{2}$ and $y_{3}$ share the same resolution and channel dimensions enables their seamless channel-wise concatenation.

This fusion produces a $(H/16) \times (W/16) \times 2M$ tensor that combines semantic latent $y_{2}$ and visual latent $y_{3}$. A subsequent residual block refines the fused representation and restores its dimensionality to $M$ channels, maintaining a compact, semantically enriched latent suitable for hyperprior analysis and entropy coding. We evaluate the effectiveness of this fusion strategy through ablation studies.

\section{The Loss Function}

Our loss function combines rate $R$ and distortion $D$ terms. The rate $R$ is associated with the expected code length of the latents $y$ and hyper-latents $z$, while the distortion $D$ measures the reconstruction fidelity between $x$ and $\hat{x}$. In this paper, we employ Mean Squared Error (MSE) as the distortion metric. A Lagrange multiplier $\lambda$ balances the trade-off between rate and distortion:

\begin{equation}\label{rate_distortion}
\begin{split}
   L &= R + \lambda D, \\
   D &= \mathbb{E}_{x \sim P_{x}}[d(x, \hat{x})],
\end{split}
\end{equation}

where $P_{x}$ denotes the distribution of natural images.

\section{Experiment}
\label{sec:experiment}

\subsection{Training Details}

Our models are trained on color PNG images from the CLIC dataset\footnote{\url{http://www.compression.cc/}} and the LIU4K dataset~\cite{LIU_dataset}. Each model targets a specific rate setting by adjusting $\lambda \in \{0.0016, 0.0032, 0.0075, 0.015, 0.03, 0.045, 0.06\}$, optimizing primarily for PSNR. The number of filters $N$ is fixed at 128. We train for 160 epochs using Adam with a batch size of 8. The learning rate is $1 \times 10^{-4}$ for the first 130 epochs and reduced to $1 \times 10^{-5}$ for the remaining 30 epochs. This training strategy ensures stable convergence and improved rate-distortion performance for all target bit rates.

\subsection{Comparisons}

\begin{figure*}[t]
\centering
\subfigure
{\begin{minipage}[t]{0.5\linewidth}
\centering
\includegraphics[scale=0.48]{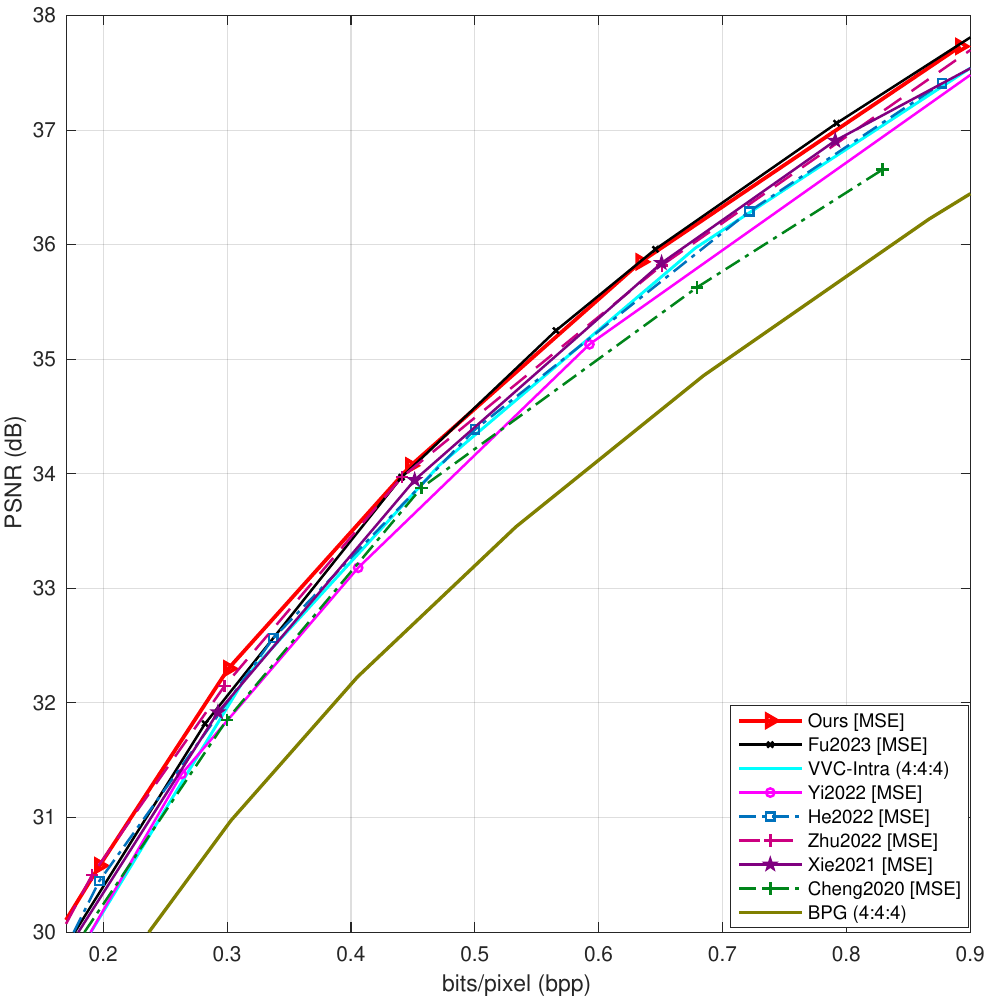}
\end{minipage}
}%
\subfigure
{\begin{minipage}[t]{0.5\linewidth}
\centering
\includegraphics[scale=0.485]{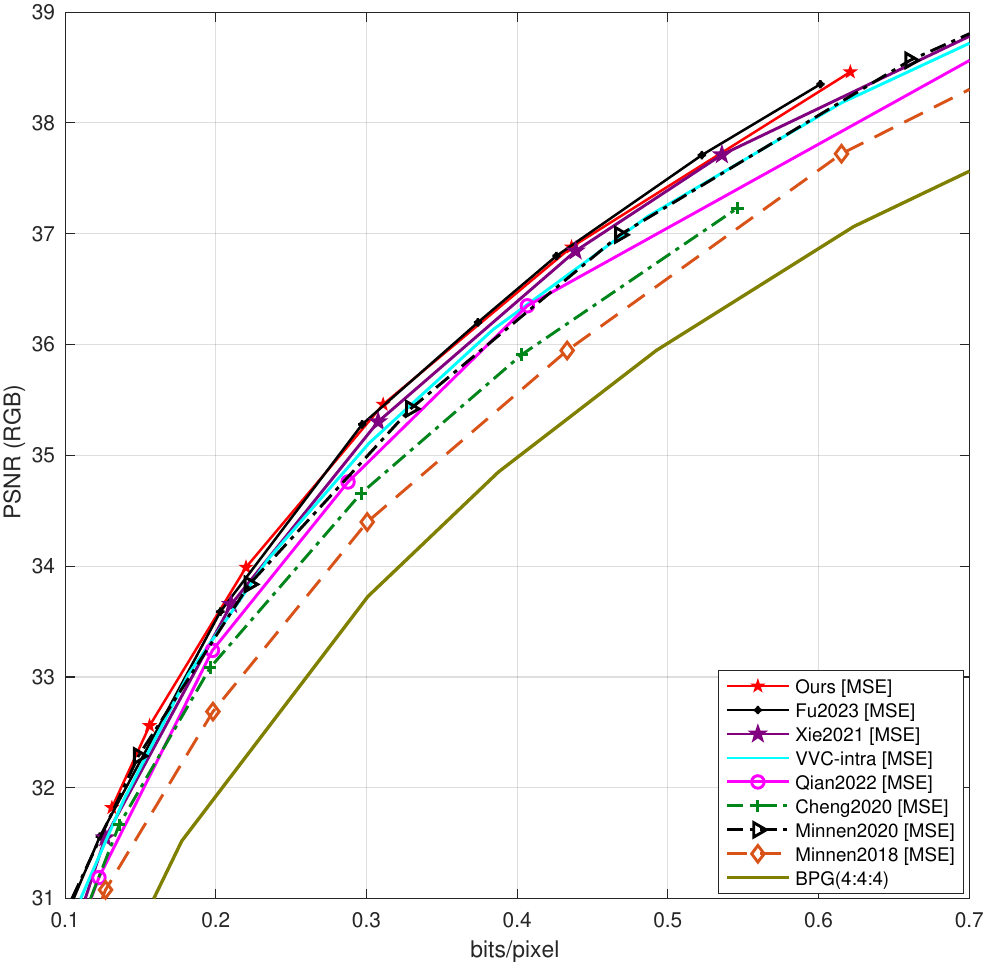}
\end{minipage}
}%
\centering
\caption{Average PSNR and MS-SSIM performance on the 24 Kodak images and 100 Tecnick images.}
\label{fig:kodak}
\end{figure*}

We compare our proposed method with recent learned compression methods and traditional codecs in terms of PSNR metric. The LIC methods include Fu2023 \cite{GLLMM},  Zhu2022 \cite{zhu2022transformerbased}, Yi2022 \cite{Qian2022_ICLR}, He2022 \cite{He_2022_CVPR},  Xie2021 \cite{xie2021enhanced}, Cheng2020 \cite{cheng2020}. The traditional methods are H.266/VVC Intra (4:4:4), and H.265/BPG Intra (4:4:4).

Fig. \ref{fig:kodak} shows the average rate-distortion curves on the 24 Kodak images and Tecnick 100 dataset. For Kodak, GLLMM achieves the top performance. Our method closely approaches GLLMM over a wide range of bit rates, outperforming other learning-based techniques and showing similar or better results than VVC (4:4:4) depending on the rate. For Tenick dataset, we remain competitive with GLLMM and surpass other learned methods.

\subsection{Performance and Speed Trade-off}

\begin{table}[!t]
\caption{Comparisons of encoding/decoding time, BD-Rate reduction over VVC, and model parameters on Kodak test set.}
\begin{center}
\begin{tabular}{ccccc}
\hline
\textbf{Methods} & \textbf{Enc. Time} & \textbf{Dec Time} & \textbf{BD-Rate} &\textbf{$\#$Params}\\ 
\hline
  VVC  & 402.3s& 0.61s& 0.0 & -\\
  
  Cheng2020 \cite{cheng2020}  &27.6s& 28.8s& 2.6 \% & 50.80 MB \\
  Hu2021 \cite{Hu_2021}  &32.7s& 77.8s& 11.1 \% & 84.60 MB \\
  He2021 \cite{He_2021_CVPR}  &20.4s& 5.2s& 8.9 \% & 46.60 MB \\ 
  Xie2021 \cite{xie2021enhanced}&4.097s &9.250s&-0.8 \% & 128.86 MB\\
  Zhu2022 \cite{zhu2022transformerbased} &0.269s &0.183s &-3.9 \% & 32.34 MB\\
  Zou2022 \cite{Zou_2022_CVPR} &0.163s &0.184s &-2.2\% & 99.86 MB\\
  Qian2022 \cite{Qian2022_ICLR}&4.78s &85.82s&3.2 \% & 128.86 MB\\
  Fu2023 \cite{GLLMM}  &420.6s& 423.8s& -3.1\% & - \\ 

\textbf{Ours} & \textbf{0.448s}  & \textbf{0.167s} & \textbf{-4.9\%} & \textbf{52.22 MB} \\

   \hline
\end{tabular}

\label{runing_time}
\end{center}
\end{table}

Table~\ref{runing_time} compares encoding/decoding times, BD-rate reductions relative to VVC~\cite{BDRate}, and model complexities for various methods on the Kodak test set. All learned approaches, including ours, were evaluated on an NVIDIA Tesla V100 GPU with 16 GB memory, whereas VVC was tested on a 2.9GHz Intel Xeon Gold 6226R CPU. Parameter counts were obtained using the PyTorch Flops Profiler, except for~\cite{GLLMM}, for which an exact measure is unavailable. Nonetheless, \cite{GLLMM} reports notably higher complexity than \cite{cheng2020}.

Some learned image compression methods~\cite{cheng2020, GLLMM, xie2021enhanced, Qian2022_ICLR} rely on serial autoregressive models that limit parallelization and thus slow down inference. By contrast, more recent approaches~\cite{zhu2022transformerbased, Zou_2022_CVPR} employ parallelizable entropy models, substantially improving speed on GPUs.

Our proposed method surpasses \cite{zhu2022transformerbased} by about 1.0\% in BD-rate reduction and achieves a total of -4.9\% relative to VVC. While our parameter count (52.22 MB) is moderately larger than that of \cite{zhu2022transformerbased} (32.34 MB) and slightly higher than \cite{cheng2020} (50.80 MB), it remains significantly smaller than other recent methods such as \cite{Zou_2022_CVPR}. Although our encoding speed is slower than those of \cite{Zou_2022_CVPR} and \cite{zhu2022transformerbased}, our decoding time is on par or faster, and both are still orders of magnitude quicker than many traditional or highly complex learned models.

Notably, our encoding time increases because we invoke pretrained image-text and text-tensor encoders (e.g., BLIP and BERT) at runtime. These encoders are employed through API calls without updating their parameters, ensuring they do not inflate our model’s parameter count. As a result, our overall parameter size remains stable, even though the initial encoding step is more time-consuming than the decoding phase.

In summary, our method balances compression performance, computational complexity, and decoding speed effectively. This trade-off makes our approach suitable for real-world scenarios, offering high-quality image compression guided by semantic information while maintaining manageable model size and practical runtime characteristics.

\subsection{Ablation Study on Semantic Integration}
\label{Ablation}

\begin{figure}[!thp] 
\centering 
\includegraphics[scale=0.5]{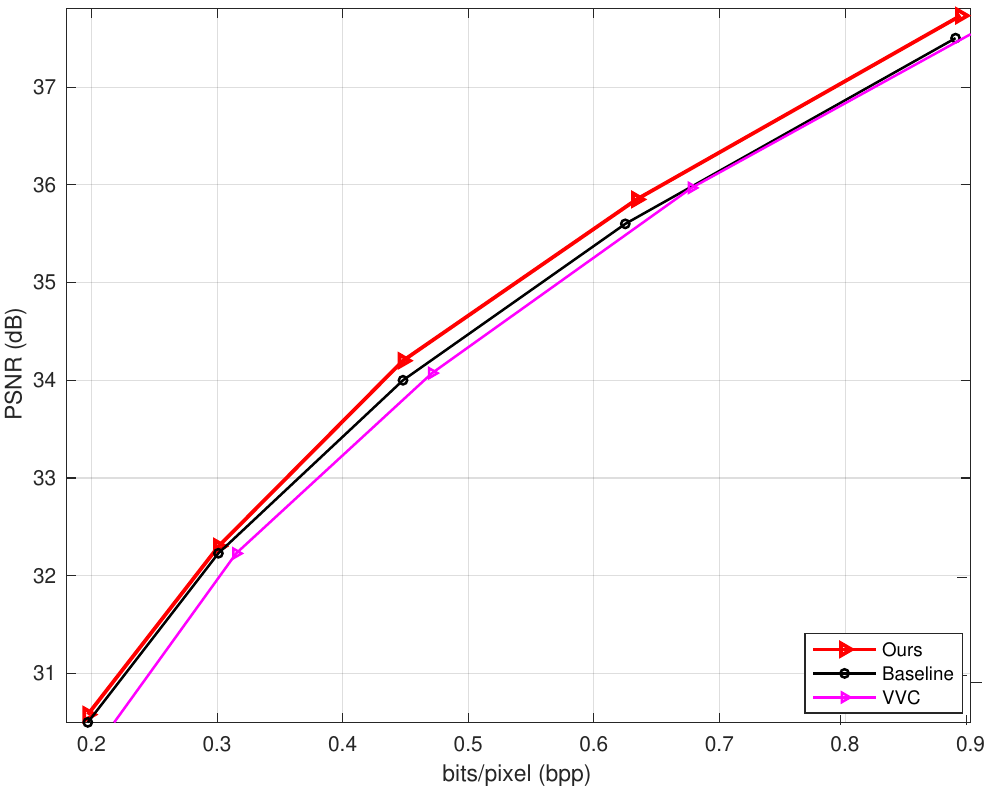} 
\caption{The impact of removing semantic-related modules from the proposed framework.} 
\label{abalation} 
\end{figure}

We conduct an ablation study by removing the semantic encoders ($g_{a1}$, $g_{a2}$) and the text-image fusion module, treating the remaining structure as our baseline. As shown in Fig. \ref{abalation},  without the integrated semantic information, the performance shows a minor decrease at low bit rates. However, at higher bit rates, the absence of textual guidance leads to a more pronounced drop of approximately 0.1~dB-0.15~dB in PSNR.

The reason is that semantic cues become increasingly beneficial as more bits are allocated to represent subtle details. At higher bit rates, the model can exploit the textual semantics to guide its feature allocation more efficiently, improving the accuracy of subtle texture and structural reconstruction. When these semantic cues are removed, the model lacks a high-level contextual reference, making it harder to optimally distribute available bits. As a result, the rate-distortion trade-off deteriorates more noticeably in these higher-rate scenarios, underscoring the value of semantic guidance in enhancing overall compression performance.

\subsection{Performance Comparison with Different Fusion Mechanisms}
\label{Ablation_spilt}

\begin{table}[!thp]
\caption{Performance with different fusion mechnisms}
\begin{center}
  \begin{tabular}{ccc}
  \hline
 \textbf{Methods} & \textbf{Bit rate}  & \textbf{PSNR } \\
  \hline
  \textbf{Element-wise Multiplication} & 0.900 & 37.16 dB   \\
  \textbf{Element-wise Addition} & 0.895 & 37.32 dB   \\
  \textbf{\textbf(ours)Channel Concatenation} & 0.890  & 37.68 dB  \\
  \hline
  \textbf{Element-wise Multiplication} & 0.203 & 30.12 dB  \\
  \textbf{Element-wise Addition} & 0.203 & 30.20 dB   \\
  \textbf{\textbf(ours)Channel Concatenation} & 0.199 & 30.61 dB \\
  \hline
\end{tabular}
\end{center}
\label{Tab:abalation_spilt}
\end{table}

Table~\ref{Tab:abalation_spilt} compares different fusion strategies for integrating image and text features at varying bit rates. Both element-wise addition and multiplication achieve suboptimal performance, as textual data is too sparse to be directly fused with image representations. This sparsity prevents effective modeling of joint dependencies, resulting in lower PSNR results.

In contrast, channel concatenation achieves consistently higher PSNR at similar or even lower bit rates. By simply merging the two feature sets along the channel dimension, it circumvents the limitations posed by sparse textual inputs. This approach preserves richer spatial information and facilitates more effective entropy modeling, ultimately enhancing overall compression quality.

\section{Conclusion}
In this paper, we introduced a semantic-enhanced learned image compression framework named \textbf{SELIC} that directly integrates high-level textual information into the compression pipeline. By embedding textual information extracted from a dedicated image-to-text encoder, our method effectively leverages semantic guidance to enhance the rate-distortion trade-off, achieving both improved fidelity and efficient decoding.

Experimental results confirm that this approach outperforms state-of-the-art LIC methods while maintaining practical computational costs. For future work, exploring more advanced image-text models could yield even more accurate semantic representations, and employing transformer-based modules for cross-modal fusion may further improve compression efficiency and quality. These directions hold promise for advancing semantic-driven learned image compression and broadening its applicability.

\bibliographystyle{IEEEbib}

\end{document}